\documentclass[10pt,a4paper]{article}
\usepackage[utf8]{inputenc}
\usepackage{url}
\usepackage{graphicx, amsmath, amsfonts,amssymb}
\usepackage{authblk}
\usepackage{color}
\pdfoutput=1
\usepackage[style=phys, sorting=none]{biblatex}
\usepackage[range-phrase = --]{siunitx}
\usepackage[font={small}]{caption}

\begin{document}

\title{\textbf{{\Large Multi-wavelength coherent diffractive imaging}}}

\author[1,*]{{\normalsize Erik Malm}}
\author[2]{{\normalsize Edwin Fohtung}}
\author[1]{{\normalsize Anders Mikkelsen}}

\affil[1]{{\small Division of Synchrotron Radiation Research, Department of Physics, Lund University, P.O. Box 118, SE-22100 Lund, Sweden}}

\affil[2]{{\small Department of Materials Science and Engineering, Rensselaer Polytechnic Institute, Troy, NY, 12180, USA.}}

\affil[*]{{\small Email: erikb.malm@gmail.com}}

\date{}

\maketitle

\begin{abstract}
Coherent diffractive imaging is a technique that recovers the sample image by numerically inverting its diffraction pattern.
We propose a generalization of this method for the inversion of multi-wavelength data.
Using this approach, we show that separate reconstructions for each wavelength can be recovered from a single polychromatic diffraction pattern.
Limitations on the number of wavelengths is provided by adapting the constraint ratio to the polychromatic situation.
The method's performance is demonstrated as a function of the source spectrum, the degree of complexity within the exit wave and the sample geometry using several numerical simulations.
Lastly, an example shows the ability to recover element-specific information using two harmonics selected from a high-order harmonic generation source.
\end{abstract}

\section{Introduction}
\begin{figure}[!ht]
	\centering
	\def\svgwidth{0.6\linewidth}
	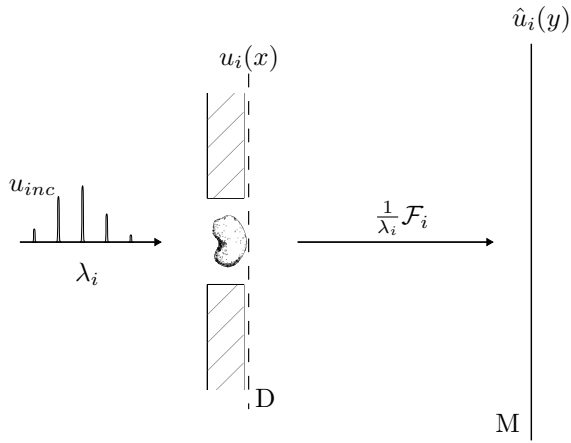
\caption{\label{fig:geometry_sketch} The experimental geometry for a multi-wavelength CDI experiment.  The incident field, comprised of several wavelengths, scatters off a localized sample producing a diffraction pattern in the measurement domain $M$.  The propagation of each field to $M$ is proportional to a wavelength-dependent Fourier transform $\mathcal{F}_i$.  The detector records an incoherent superposition of diffraction patterns according to Eq.~\ref{eq:intensity}.}
\end{figure}

Coherent lensless imaging is a collection of techniques which replace the optics within a conventional microscope with a numerical algorithm.
These techniques are attractive and applicable in areas where it is challenging to design and fabricate high quality diffractive and refractive optics via nano-fabrication methods.
In light-based microscopy, these areas are primarily in regimes with wavelengths below the visible spectrum.
Coherent diffractive imaging (CDI) is one such technique.
The experimental geometry for plane-wave CDI is shown in Fig.~\ref{fig:geometry_sketch}.

CDI uses far-field diffraction measurements and prior knowledge to recover the sample image.
Numerically, this is accomplished by recovering the phase in the diffraction data.
Phase retrieval algorithms impose constraints on the field estimate through a series of projections which are applied in an iterative fashion \cite{Marchesini2007}.
An estimate of the field is obtained by alternating between these projections in real (sample-plane) and Fourier (detector-plane) space.
Back-propagation of the field from the detector to the sample-plane results in the sought after exit surface wave.
Once a solution has been found, a forward model can be used to relate the exit surface wave to certain sample properties such as electron and spin densities \cite{Tripathi2011, Kfir2017, Rose2018} or crystal strain \cite{Robinson2009, Karpov2019}.

CDI typically relies on coherent light to exploit the Fourier relationship between the exit wave and its far field; however, short-wavelength sources such as high-order harmonic generation or undulator-based sources contain several harmonics \cite{Heyl2016, Inoue2018}.
As a result, monochromators are necessary to attenuate these harmonics to obtain fully coherent diffraction data.
Higher beamline throughput and additional sample information can be obtained if these harmonics are properly utilized.

Complementary coherent lensless techniques which include a two-pulse CDI scheme, Fourier transform holography, coherent modulation imaging and ptychography have been demonstrated previously on polychromatic data.
The two-pulse scheme utilizes a varying time delay between the pulses to recover the sample image \cite{Witte2014}.
Coherent modulus imaging is capable of single-shot polychromatic imaging through the use of a known phase plate \cite{Dong2018}.
Ptychography uses the redundancy in the data to recover spectral information \cite{Batey2014, Thibault2013}, while Fourier transform holography relies on the spatial separation of the autocorrelation functions \cite{Willems2017}.
A comparison of these techniques is beyond the scope of this work; overviews of these techniques are given in \cite{Chapman2010,Paganin2006}.

In addition, polychromatic CDI experiments have been performed with the aim of increasing beamline efficiency and reducing exposure times \cite{Chen2009,Dilanian2009,Abbey2011,Chen2012,Malm2020}.
These experiments were an important step in demonstrating that the source's temporal coherence can be relaxed.
A drawback of these methods is the added algorithmic complexity and additional computational resources needed for the propagation of polychromatic fields.
Additionally, these methods recover a single exit surface wave or transmission function making them ill-suited for the recovery of spectral information from general samples.
For these reasons, previous methods are restricted to a subset of possible samples.

%	New approach
The proposed method lifts these restrictions on the sample by recovering an exit surface wave for each wavelength within the data.
This is accomplished by allowing each reconstruction to evolve separately.  
The reconstructions are not completely independent, but are coupled through their common support shape, spectral weights and diffraction measurements.  
The method is capable of capturing the spectral response from complex samples as if several separate monochromatic CDI experiments were performed with different wavelengths.  
While this method is described for multi-wavelength CDI, it is applicable to situations where an incoherent superposition of signals is present.
This situation arises in several contexts; it occurs when the field contains different polarization states \cite{Eisebitt2003,Eisebitt2004}, time delays beyond the source's longitudinal coherence, or for multiple wavelength sources.
These situations occur in  pump-probe experiments, for samples containing charge and magnetic contributions, or for polychromatic sources.
A detailed description of this last situation is provided in the following treatment.

\section{Theory}
%	Forward model
The experimental geometry, shown in Fig.~\ref{fig:geometry_sketch}, is considered where the incident field, comprised of multiple wavelengths, $\lambda_i$, interacts with a spatially localized sample to produce a polychromatic diffraction pattern on the detector.
Each field, $u_i$, associated with $\lambda_i$, is treated as fully coherent.
The sample is localized spatially and treated within the projection approximation effectively making it two-dimensional.
The sample support, $\Omega$, is contained in a fixed domain $D \subset \mathbb{R}^2$ and the diffraction patten is collected in the detector domain $ M\subset \mathbb{R}^2$ which is placed in the far-field.
The propagation of the field to the detector plane is given by the Fraunhofer equation:
\begin{align} \label{eq:far_field}
\hat{u}_i(y) \propto \frac{1}{\lambda_i} \mathcal{F} \{ u_i(x) \} (q_i) \biggr\rvert_{q_i = \frac{y}{\lambda_i z}} ,
\end{align}
where $\mathcal{F}$ is the Fourier transform operator and $x$, $q$ are sample-plane coordinates and the spatial frequencies respectively.
The intensity from each $\lambda_i$ is given by $I_i(y) = \vert \hat{u}_i (y) \vert^2$.  
When each $\lambda_i$ is sufficiently separated from its neighboring wavelengths, the interference between fields can be neglected \cite{Dilanian2009, Quiney2010}.
This occurs when the detector integration time is long compared to the beat frequency.
In this situation, the total intensity, $I(y)$, on the detector is described by an incoherent sum of far-field intensities
\begin{equation}
I(y)  = \sum_{i=1}^P I_i(y) ,
\label{eq:intensity}
\end{equation}
where $y \in M$ is the detector-plane coordinate and $\lambda_i$ is sorted in increasing order such that $\lambda_1 < \lambda_2 < \dots\lambda_P$.

The numerical procedure for calculating $I(y)$ depends on the discretization procedure, which in turn, specifies the numerical approach used to propagate $u_i$ between $D$ and $M$.
There are two possible approaches: The spatial coordinate, $x$, could be discretized on the same grid for each $u_i$ leading to a wavelength-dependent propagation operator $\mathcal{F}_i$.
The alternative is to use a wavelength-independent $\mathcal{F}$ (fast Fourier transform) with a wavelength-dependent grid for $x$.
The latter is preferable for two reasons:  First, a wavelength-independent $\mathcal{F}$ avoids interpolation operations making it faster to compute while a wavelength-dependent grid leads to different support constraints for each field which allows each one to be reconstructed separately.
To facilitate this approach we define a new field variable 
\begin{align} \label{eq:u_check}
\check{u}_i \left(x \right) := \frac{\lambda_i}{\lambda_0} \, u_i \left(\frac{\lambda_i}{\lambda_0} x \right) 
\end{align}
where $0$ corresponds to a reference index.  
Now, the far field, $\hat{u}_i$, and $\check{u}_i$ are related through a Fourier transform.
The details of this approach together with uniqueness considerations are provided next.

For proper sampling of the diffraction pattern in a CDI experiment, the scattered field must originate from a small localized region in space \cite{Bates1982,Miao1998}.
This can be accomplished if the sample is small and isolated or placed within an opening of an opaque mask.
Physically, each field has the same support shape and size given by either the shape of the sample itself or the mask opening.
Numerically, however, the support size depends on the real-space pixel size, which, depends on the wavelength.
The dependence of the sample-plane pixel size, $\Delta_s^i$, on the detector pixel size, $\Delta_d$, the sample-to-detector distance, $z$ and number of pixels along one direction, $N$ is given by
\begin{equation}
\Delta_s^i = \frac{\lambda_i z}{\Delta_d N} .
\label{eq:pixel_size}
\end{equation}
This relation is obtained through a comparison between the analytic equation for far-field propagation and the discrete fast Fourier transform used for the numerical implementation.  
This relationship leads to a wavelength-dependent scaling of the support domains in discrete space described by
\begin{equation}
	\Omega_i = \varphi_i(\Omega_0) ,
	\label{eq:support_scaling}
\end{equation}
where $\Omega_i$ and $\Omega_0$ are the supports for $\check{u}_i$ and $\check{u}_0$ respectively.  
Here, $\check{u}_0$ and $\Omega_0$ denote the field and support associated with the largest spectral intensity, $S_0$.
The transformation has the form: $\varphi_i(x) = \frac{\lambda_i}{\lambda_0} x + \delta_i$, where $\delta_i$ can be set to zero due to the translational invariance of Eq.~\ref{eq:intensity}.  
The scaling of the supports, described by Eq.~\ref{eq:support_scaling}, lead to separate support constraints which ultimately allows this approach to recover each $\check{u}_i$ separately.

%	Algorithm projections/details
Based on the previous discussion a set of projections are defined which form the basis for polychromatic phase retrieval.
The projections are determined by the source spectrum, $S_i$, and the measured polychromatic far-field diffraction intensities, $I_m$.  
The support constraint can be determined throughout the iterative process similar to the procedure which is described in \cite{Marchesini2003}.  
Each $\check{u}_i$ must fulfill a separate support constraint described by Eq.~\ref{eq:support_scaling}.  This constraint is fulfilled through the application of the support projection separately to each $\check{u}_i$.
The support projection becomes
\begin{equation}
\Pi_{S} \check{u}_i(x) = 
	\begin{cases}
	\check{u}_i(x) & x \in \Omega_i \\
	0 & \text{otherwise}
	\end{cases},
\label{eq:support_projection}
\end{equation}
where each $\Omega_i$ satisfies the relationship from Eq.~\ref{eq:support_scaling}.
Next, each $\check{u}_i$ is normalized to the measured polychromatic amplitude data, $g(y) = \sqrt{I_m(y)}$, according to
\begin{equation}
\Pi_N \check{u}_i(x) =  \sqrt{S_i} \; \check{u}_i(x) \; \frac{ \Vert g \Vert }{\Vert \check{u}_i \Vert}  ,
\label{eq:spectral_projection}
\end{equation}
where $\Vert \cdot \Vert = \sqrt{\int \lbrace \cdot \rbrace ^2} $ indicates the usual $L^2$ norm.
The spectral intensities, $S_i$, are a measure of the relative contribution from each wavelength to the total intensity.
The operation defined in Eq.~\ref{eq:spectral_projection} ensures the calculated intensity, $I(y)$, has the correct spectral weightings.
In Fourier space, the magnitude projection is given by
\begin{equation}
\tilde{\Pi}_M \hat{u}_i(y) = \frac{\hat{u}_i(y)}{\sqrt{I(y)} + \epsilon} g(y)
\label{eq:modulus_projection}
\end{equation}
where $\epsilon$ is a small positive constant.  This projection rescales the magnitude of each $\hat{u}_i$ corresponding to a projection onto a $(P-1)$-dimensional sphere, where $P$ is the number of wavelengths present in the diffraction data (Eq.~\ref{eq:intensity}). $\epsilon$ is included in the denominator to prevent issues for $\sqrt{I(y)}$ values which are close to zero.
Next, we provide a discussion on the uniqueness of the solution and its dependence on the spectrum and support geometry.

%Uniqueness: Constraint ratio 
Uniqueness is an important aspect of any inverse problem and is discussed through the adaptation of the concepts in \cite{Elser:2007} to the multi-wavelength situation.
By uniqueness, it is meant a unique equivalence class consisting of solutions related through translation, inversion and global phase rotation, $\lbrack u \rbrack = \lbrace u \in H(D)  : u(x) \sim u(x+a), u(x) \sim u^*(-x), u(x) \sim e^{i \alpha} u(x) \rbrace$, where $\alpha \in \mathbb{R}$, $a \in \mathbb{R}^2$, $u^*$ is the complex conjugate of $u$ and $H$ is a suitable function space in $D$.

Uniqueness was characterized previously through the constraint ratio which measures the number of constraints relative to the number of unknowns through
\begin{equation}
\Sigma = \frac{\textrm{number of independent autocorrelation coefficients}}{\textrm{number of independent object coefficients}}
\end{equation}
with $\Sigma > 1$ being a necessary condition for uniqueness.
A unique solution becomes more likely for larger $\Sigma$ values.
Due to the centro-symmetry of the autocorrelation function, only half the coefficients are independent which leads to
\begin{equation}
\Sigma_{\textrm{mono}} = \frac{\vert A \vert}{2 \vert \Omega \vert}
\label{eq:mono_constraint_ratio}
\end{equation}
for the monochromatic case, where $\vert A \vert$ and $\vert \Omega \vert$ are the size of the autocorrelation and sample supports respectively.
In the polychromatic case, $\Sigma$ must include the total size of the autocorrelation function as well as the total number of coefficients within all of the supports.
Through a direct application of these ideas, the polychromatic constraint ratio becomes
\begin{equation}
\Sigma = \frac{\lvert \bigcup\limits_{i=1}^P A_i \rvert}{2 \sum\limits_{i=1}^{P} \vert \Omega_i \vert}
\label{eq:poly_constraint_ratio}
\end{equation}
where $A_i$ is the autocorrelation support associated with $u_i$.

For convex $\Omega$, $\lvert \bigcup\limits_{i=1}^P A_i \rvert = \lvert A_1 \rvert$ and it is evident that increasing $P$ significantly reduces $\Sigma$.
This can make the solution more difficult to find or potentially non-unique.
This can be compensated for by using a non-convex or multiply connected support geometry.
For a multiply connected support each $A_i$ become partially separated with respect to each other increasing $\lvert \bigcup\limits_{i=1}^P A_i \rvert$ and resulting in larger $\Sigma$ values.
Ultimately, having a large discrepancy between the wavelengths and separated support components will strengthen the projection from Eq.~\ref{eq:support_projection} and increase $\Sigma$ resulting in improved algorithm performance.
These concepts are illustrated in the next section through a series of simulations with different support shapes and varying wavelength separations.

\section{Numerical simulations}
\begin{figure}[!htbp]
	\centering
	\includegraphics[width=0.55\columnwidth]{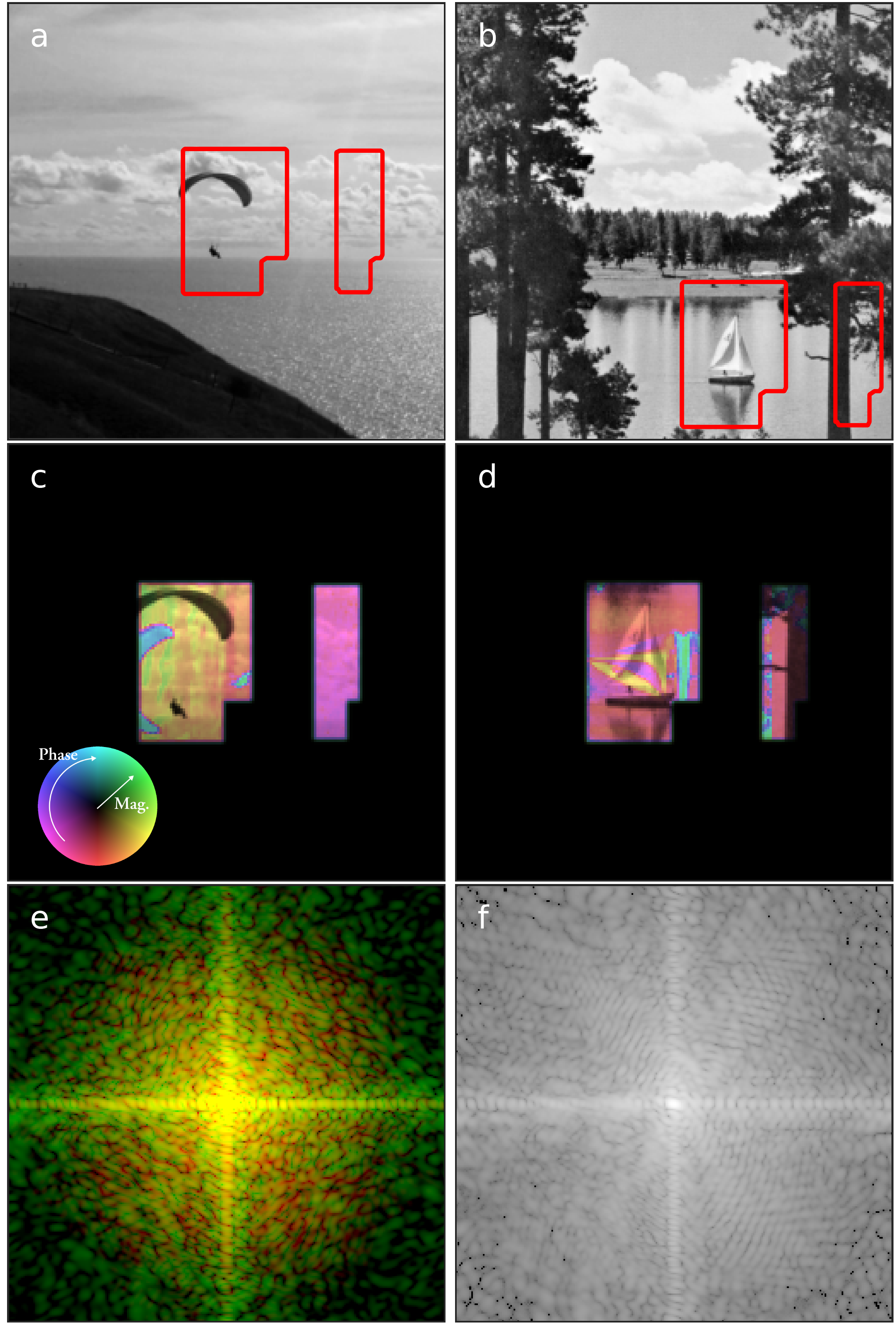}
	\caption{\label{fig:para_lake_and_diffraction} (a,b) The images used to create the exit waves for $\lambda_1$ and $\lambda_2$.  The sample support boundaries are indicated by the red lines.  (c,d) The complex-valued exit waves created using the images from (a,b) for the amplitude and rotated versions for the phase distribution.  (e) The polychromatic diffraction pattern showing the individual contributions from $\lambda_1$ and $\lambda_2$ in red and green.  (f) The polychromatic data used by the phase retrieval algorithm which consists of an incoherent sum of intensities.}
\end{figure}

Results from several numerical simulations are provided to illustrate the algorithm's performance.
The first set of simulations uses the images from Fig.~\ref{fig:para_lake_and_diffraction}(a,b) to construct exit waves associated with wavelengths $\lambda_1$ and $\lambda_2$.
These simulations explore the performance as a function of the source spectrum ($\lambda_1 / \lambda_2$), the support shape, and degree of complexity, $\phi_{\textrm{max}}$, in the exit surface waves.
In the second simulation, a multi-wavelength diffraction pattern using two harmonics taken from a high-order harmonic generation source were used to determine the elemental distribution of a sample consisting of a \SI{100}{nm} Al thin film with \SI{40}{nm} Si inclusions.
This simulation shows that the algorithm is capable of handling uncertainty in the diffraction pattern's spectral content which arises from sample absorption.

\begin{figure}[!ht]
	\centering
	\includegraphics[width=0.75\linewidth]{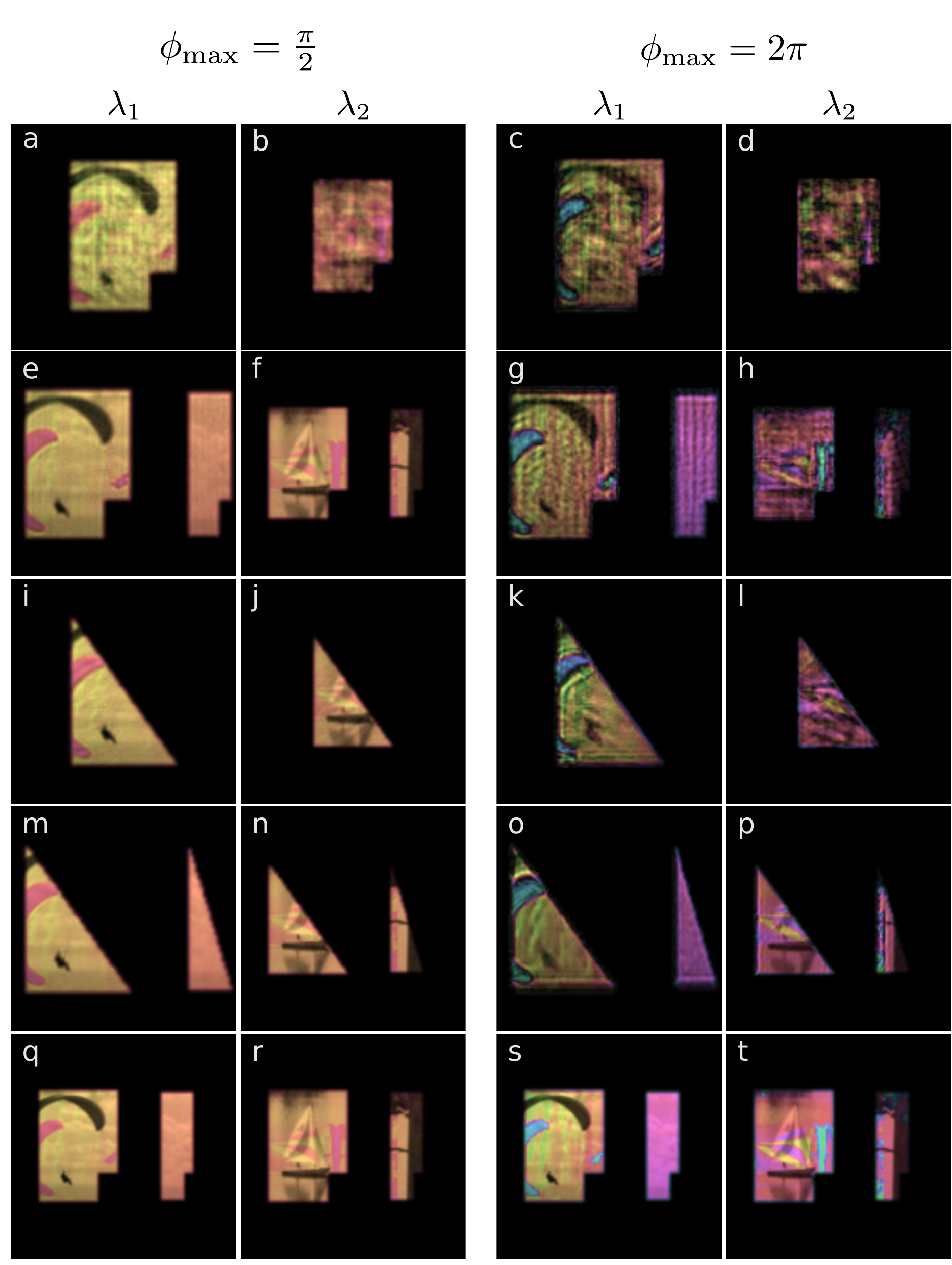}
	\caption{\label{fig:reconstructions} Reconstructions for different support shapes and degrees of complexity for both wavelengths ($\lambda_1$, $\lambda_2$).  Each row corresponds to a different support shape where the last row shows the true exit waves.  The leftmost two columns correspond to $\phi_{\textrm{max}} = \pi/ 2$ while the two columns on the right show reconstructions for $\phi_{\textrm{max}} = 2 \pi$. The magnitude and phase of the ESWs are mapped to the value and hue of the image using the colormap which is shown in Fig.~\ref{fig:para_lake_and_diffraction}(c).}
\end{figure}

The first set of simulations utilizes the images from Fig.~\ref{fig:para_lake_and_diffraction}(a,b) to make the amplitude distribution for the exit waves and rotated versions for the phase distribution.
Four different support shapes were used consisting of one or two openings which are either rectangular or triangular in shape.
The reconstructions shown in Fig.~\ref{fig:reconstructions} show these different support shapes.
The support shapes consisting of one rectangle, two rectangles, one triangle and two triangles that have constraint ratios: $\Sigma_{\textrm{mono}} = \{ 2.02, 2.78, 2.73, 4.21 \}$ respectively.
As we will see later in this section these values significantly affect the algorithm performance.
The degree of difficultly also depends on the range of phase values (degree of complexity) in the exit waves.
Sample exit waves with $\phi_{\textrm{max}} \ in \{ 0, \, \pi / 2, \, \pi, \, 2 \pi \}$ were used to explore this dependence.
Two exit waves, corresponding to $u_1(x)$ and $u_2(x)$, with $\phi_{\textrm{max}} = 2 \pi$ are shown in Fig.~\ref{fig:para_lake_and_diffraction}(c,d).
Complex-valued images are visualized by mapping the intensity and phase of $u_i(x)$ to the value and hue of the image respectively.
The inset in Fig.~\ref{fig:para_lake_and_diffraction}(c) illustrates this mapping.

Once the $u_i(x)$ have been created the waves are propagated to the far-field and interpolated to the fixed grid spacing of the detector.
In this simulation, two waves are present at the detector; the contribution from each $\hat{u}_i(y)$ is shown in Fig.~\ref{fig:para_lake_and_diffraction}(e) using red and green for the different wavelength contributions.
Finally, an incoherent superposition of intensities are calculated according to Eq.~\ref{eq:intensity}.
Noise was introduced into the data by including Gaussian white noise and through the quantization of the intensity values.
The signal-to-noise ratio was set to 60dB which was calculated according to: $SNR = 10 \log {\parallel I_{exact} \parallel} / {\parallel I_m - I_{exact} \parallel}$, where $I_{exact}$ and $I_m$ correspond to the ideal and corrupted intensities respectively.  
An example of the measured polychromatic data is shown in Fig.~\ref{fig:para_lake_and_diffraction}(f).

The reconstruction procedure follows the same approach taken by monochromatic phase retrieval algorithms; this consists of projecting the estimate between constraint sets in an alternating fashion until a solution is found.
Phase retrieval was performed by alternating between the error reduction and hybrid input-output methods \cite{Fienup1982} in an alternating fashion where the monochromatic projections were exchanged with their polychromatic counterparts described by Eqs.~\ref{eq:support_projection} - \ref{eq:modulus_projection}.

The algorithm performance is isolated from any support determination scheme by using a fixed support.
The final simulation will determine the support iteratively using the first approach described next.
We identify two approaches which could be used to determine the sample support.
First, one could determine $\Omega_0$ by thresholding a Gaussian smoothed version of $\vert u_0 \vert $, then resize $\Omega_0$ according to Eq.~\ref{eq:support_scaling} to obtain the remaining $\Omega_i$.
This procedure works well as long as the sample does not exhibit significant wavelength-dependent absorption.
Otherwise, $\Omega_0$ could acquire internal structure which may be absent in the other $\Omega_i$ resulting in a degradation of the reconstruction quality for the remaining $u_i$.  This can be alleviated by using the union of the supports by first determining $\Omega_i$ by thresholding all $\vert u_i \vert $ individually, applying $\varphi_i^{-1}$ to account for the different pixel sizes, taking the union of the supports and then rescaling with $\varphi_i$ to obtain each $\Omega_i$.

%Results discussion:
Several of the reconstructions are shown in Fig.~\ref{fig:reconstructions} for the four supports and for two different degrees of complexity ($\phi_{\textrm{max}} \in \{\pi / 2,\ , 2\pi \}$).
The reconstructions in Fig.~\ref{fig:reconstructions} all have $\lambda_1 / \lambda_2 = 0.75$.
The wavelength dependence of the pixel size (Eq.~\ref{eq:pixel_size}) is clearly visible in the reconstructions which make $u_1$ appear larger than $u_2$.
The reconstructions with a triangular support (Fig.~\ref{fig:reconstructions}(i-l)) show a clear improvement over the reconstruction with a rectangular support (Fig.~\ref{fig:reconstructions}(a-d)).
The difference can be explained through $\Sigma_{\textrm{mono}}$ which corresponds to $2.02$ and $2.73$ for the rectangular and triangular supports respectively.
In addition, the error increases for samples with higher degrees of complexity which is clear by comparing the reconstructions of the first and last two columns in Fig.~\ref{fig:reconstructions}.

\begin{figure}[!ht]
	\centering
	\includegraphics[width=0.6\columnwidth]{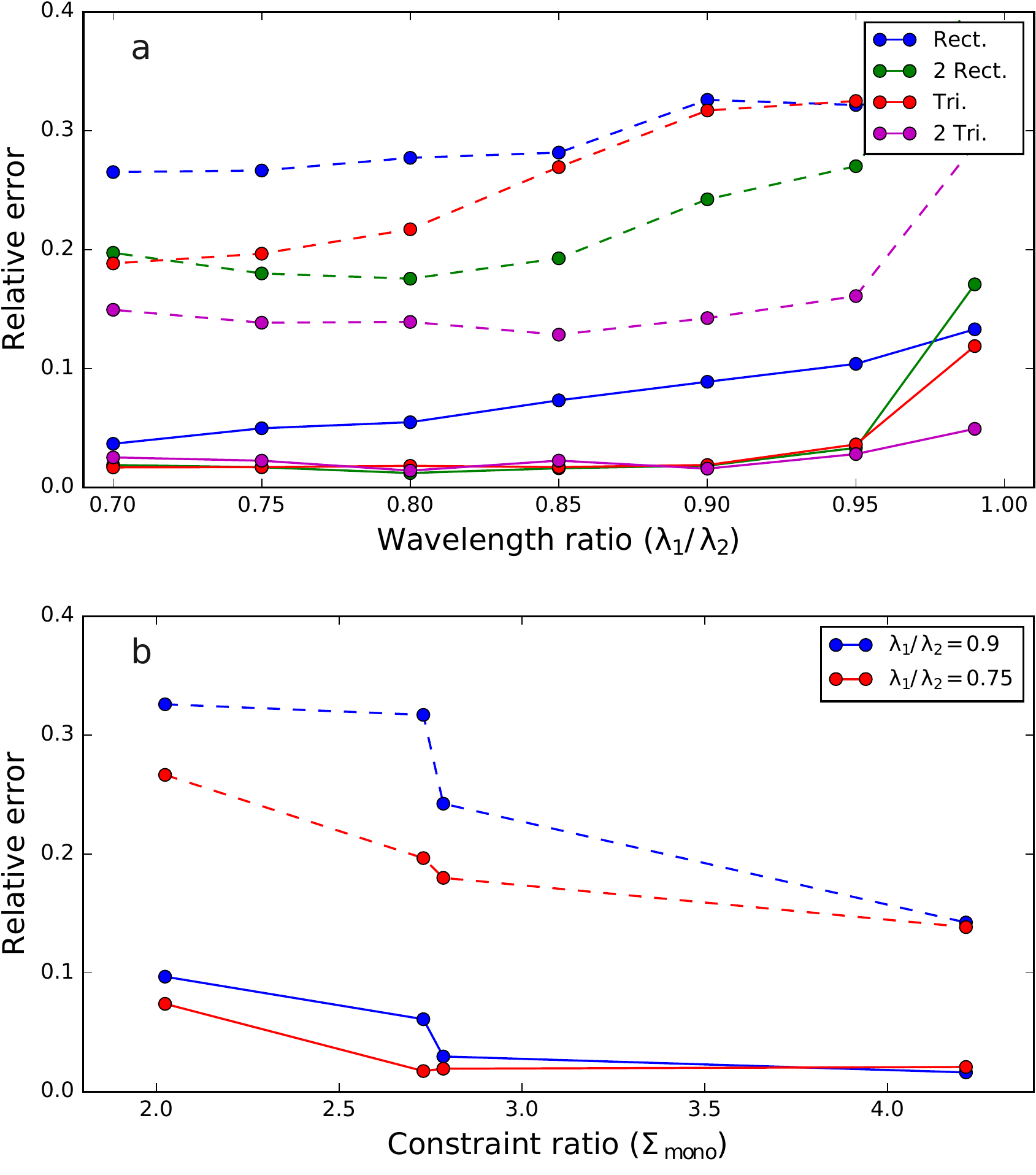}
	\caption{\label{fig:errors} (a) The relative error as a function of $\lambda_1 / \lambda_2$ for different support geometries.  The solid and dashed lines correspond to $\phi_{\textrm{max}} =0$ and $\phi_{\textrm{max}} =2 \pi$ respectively. (b) The relative error as a function of the constraint ratio.  The four constraint ratios correspond to the four different sample supports.  The solid and dashed lines correspond to $\phi_{\textrm{max}} =\pi / 2$ and $\phi_{\textrm{max}} =2 \pi$ respectively.  Each point corresponds to the average error from 20 independent trials.}
\end{figure}

These observations are supported by the error plots in Fig.~\ref{fig:errors}.
The relative error was calculated relative to the known images using the equation: $E = \sum_{i=1}^P {\parallel \alpha \vert u_i \vert - \vert u_i^{exact} \vert \parallel} / {\parallel u_i^{exact} \parallel}$, where $\alpha = {\parallel u_i^{exact} \parallel} / {\parallel u_i \parallel}$ is a normalization factor.
The absolute values were compared to avoid any issues associated with global phase offsets between the two images.
Because the supports were several pixels larger than the exact supports, each $u_i$ was registered to $u_{exact}$ before computing the error.
The registration procedure is described in \cite{Guizar-Sicairos2008}.
Each point in Fig.~\ref{fig:errors} represents the average error from 20 independent reconstructions.
The reconstruction procedure was terminated after 2000 iterations.

Fig.~\ref{fig:errors}(a) shows the errors as a function of $\lambda_1 / \lambda_2$ for $\phi_{\textrm{max}} \in \{0, \, 2\pi \}$.
The solid and dashed lines correspond to $\phi_{\textrm{max}} =0 $ and $\phi_{\textrm{max}} = 2 \pi$ respectively.
The error clearly decreases as $\lambda_1 / \lambda_2$ becomes smaller.
This results in the numerator in Eq.~\ref{eq:poly_constraint_ratio} growing faster with $\lambda_1 / \lambda_2$ than the denominator resulting in larger $\Sigma$ values.
Out of the four supports it is apparent that the support consisting of two triangular openings provides the best reconstructions.
The effect of $\Sigma$ on the reconstruction quality is shown in Fig.~\ref{fig:errors}(b).
The solid and dashed lines correspond to $\phi_{\textrm{max}} = \pi / 2$ and $\phi_{\textrm{max}} = 2 \pi $ respectively.
%Abrupt change at ...

\begin{figure*}[!t]
	\centering
	\includegraphics[width=\linewidth]{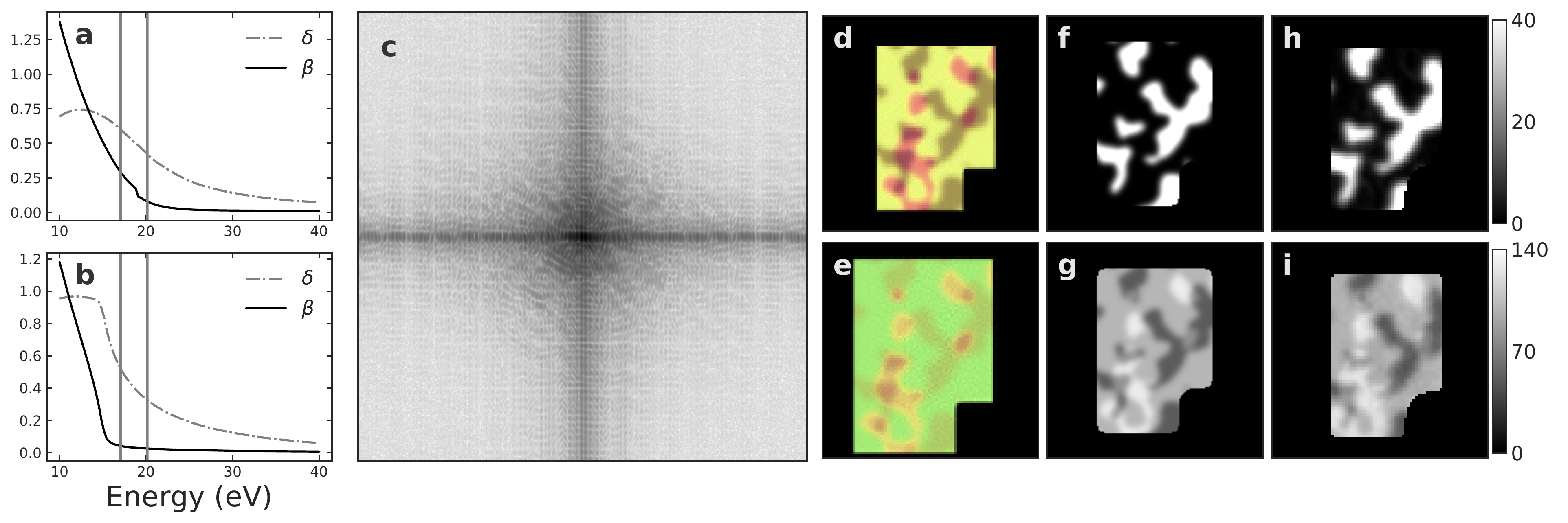}
	\caption{\label{fig:al_si} The recovery of the spatial distributions of aluminum and silicon within the sample from a single polychromatic diffraction pattern.  The refractive indices for Si (a) and Al (b) as a function of photon energy with $\lambda_1$ and $\lambda_2$ indicated by the vertical lines. (c) The polychromatic diffraction pattern shown in log scale.  The sample consists of a thin film of Al (g) with Si inclusions (f) and comprised of an opaque gold mask which defines the support. (d,e) The recovered exit surface waves for both wavelengths.  The complex-valued images are shown by mapping the amplitude and phase to value and hue respectively using the same colormap which is shown in Fig.~\ref{fig:para_lake_and_diffraction}(c).  The recovered Si (h) and Al (i) thicknesses in units of nanometers.}
\end{figure*}

% Al with Si inclusions simulation
The final simulation illustrates how two harmonics from a high-order harmonic generation source can be utilized to extract the elemental spatial distribution within a sample.
The sample consists of Si and Al with spatial distributions which are shown in Fig.~\ref{fig:al_si}(f,g) respectively.
We assert that the experimental geometry (sample-detector distance, pixel size, etc.) and source spectrum have been measured and a known quantities.
The $11^\textrm{th}$ and $13^\textrm{th}$ harmonics from a high-order harmonic generation source are used to create the diffraction data (Fig.~\ref{fig:al_si}(c)).
The selection of these two harmonics can be accomplished using the spectral filtering from multilayer mirrors.
The exit surface waves were created through the transmission of two plane-waves through a sample consisting of Au, Al and Si using the projection approximation.
$\delta$ and $\beta$ of the refractive indices ($n=1-\delta +i \beta$) for Si and Al are shown as a function of photon energy in the plots in Fig.~\ref{fig:al_si}(a,b), where $\lambda_1$ and $\lambda_2$ are indicated by vertical lines.
The values for $\delta$ and $\beta$ were obtained from previously tabulated values \cite{Edwards1985,Smith1985}.
A mask consisting of \SI{1}{\micro m} thick Au was used to define the sample support (shown in Fig.~\ref{fig:para_lake_and_diffraction} in red).
A \SI{100}{nm} Al thin film with \SI{40}{nm} Si inclusions cover the opening on the left while the other opening was uncovered.
The uncovered opening plays a dual role: Its presence increases $\Sigma$ and thereby improves the reconstruction quality, but also allows for the determination of the incident field.
The incident field is used to account for any amplitude or global phase rotations between the two reconstructions.
%We account for any global phase rotation between the two reconstructions by setting the waves equal to one within the window.
%These waves are then input into a linear solver to calculate the thickness distributions.

%Reconstruction procedure
The reconstruction procedure was similar to the previous simulations.
Unlike these simulations, the support was determined iteratively by thresholding a Gaussian filtered $\vert u_0 \vert $.
In addition, the reconstruction quality was improved by running several independent trials and combining them in a guided approach similar to the procedure described in \cite{Chen2007}.
The best solutions have both the lowest error to the data and the smallest support size.
These solutions constitute a subset which are non-dominated by any other solution (constituting the Pareto front).
This method allowed for a guided approach while also updating the supports.

An issue arises as the sample absorption alters the spectral content of the incident field.
The sample absorption changed the spectral intensities from $\{0.6, 0.4\}$ before the sample to $\{0.586, 0.414\}$ at the exit surface.
Using the wrong $S_i$ values can result in artifacts within the reconstructions.
These artifacts were attenuated by removing the spectral constraint near the end of the phase retrieval process.
In this case, the artifacts were quite small, but we have noticed that the algorithm is capable of handling larger uncertainties using this procedure.
The recovered exit waves are shown in Fig.~\ref{fig:al_si}(d,e).
Finally, a linear solver was used which uses the absorption difference between $\lambda_1$ and $\lambda_2$ to determine the thickness distributions of Si and Al (Fig.~\ref{fig:al_si}(j,k)).
Prior to solving for the distributions, the exit waves were convolved with a Gaussian to attenuate the high-frequency noise which is present in the reconstructions.
This results in a small blurring and loss of spatial resolution which can be seen in Fig.~\ref{fig:al_si}(j,k).
The recovered distributions of Si and Al show the method is capable of recovering element-specific sample information even when the $S_i$ values are not known exactly.

%Overall discussion

\section{Conclusions}

In conclusion, we have proposed and investigated a general algorithm capable of recovering separate exit surface waves for each wavelength in a polychromatic diffraction pattern.
In this method, propagation is performed using a wavelength-independent fast Fourier transform requiring no interpolation operations which results in simple and fast propagation calculations.
A critical aspect of the algorithm stems from the observation that the sample pixel size contains a wavelength dependence.
This is exploited to obtain different support constraints for each field which allows each field to evolve and be reconstructed separately.
It is shown through numerical simulations that the approach can recover element-specific information and is able to cope with uncertainty within the spectrum.
The introduction of additional unknowns results in a lower constraint ratio, which in effect, places limitations on the number of wavelengths which can be utilized.
This, however, can be compensated for through the design of the sample mask.
The additional information recovered through this method gives a distinct advantage to polychromatic sources for CDI experiments.

\section*{Acknowledgments}
This work was supported by the U.S. Department of Defense (DOD) Air Force Office of Scientific Research under Award No. FA9550-18-1-0196, the NanoLund Center for Nanoscience at Lund University and the Swedish Research Council.

\printbibliography

\end{document}